\documentclass[12pt]{article}

\usepackage{epsfig}
\usepackage{amsfonts}
\usepackage{amssymb,amsmath}

\begin{document}

\def\identity{{\rlap{1} \hskip 1.6pt \hbox{1}}}
\def\half{{\textstyle{1\over2}}} 

\newcommand{\NP}{{\em Nucl.\ Phys.\ }}
\newcommand{\AP}{{\em Ann.\ Phys.\ }}
\newcommand{\PL}{{\em Phys.\ Lett.\ }}
\newcommand{\PR}{{\em Phys.\ Rev.\ }}
\newcommand{\PRL}{{\em Phys.\ Rev.\ Lett.\ }}
\newcommand{\PRP}{{\em Phys.\ Rep.\ }}
\newcommand{\CMP}{{\em Comm.\ Math.\ Phys.\ }}
\newcommand{\MPL}{{\em Mod.\ Phys.\ Lett.\ }}
\newcommand{\IJMP}{{\em Int.\ J.\ Mod.\ Phys.\ }}

\pagestyle{plain}
\setcounter{page}{1}

\baselineskip16pt

\begin{titlepage}

\begin{flushright}
MIT-CTP-3489\\
hep-th/0404102
\end{flushright}
\vspace{13 mm}

\begin{center}

{\Large \bf Perturbative computations in string field theory}\\
\vspace{3mm}
Material from lectures presented at TASI 2003, Boulder, Colorado;\\
Supplement to TASI 2001 lectures by Taylor and Zwiebach

\end{center}

\vspace{7 mm}

\begin{center}

Washington Taylor\\

\vspace{3mm}
{\small \sl Center for Theoretical Physics} \\
{\small \sl MIT, Bldg.  6} \\
{\small \sl Cambridge, MA 02139, U.S.A.} \\
{\small \tt wati {\rm at} mit.edu}\\
\end{center}

\vspace{8 mm}

\begin{abstract}
These notes describe how perturbative on-shell and off-shell
string amplitudes can be computed using string field theory.
Computational methods for approximating arbitrary amplitudes are
discussed, and compared with standard world-sheet methods for
computing on-shell amplitudes.  These lecture notes are not
self-contained; they contain the material from W.\ Taylor's TASI 2003
lectures not covered in the recently published ``TASI 2001'' notes
{\tt hep-th/0311017} by Taylor and Zwiebach, and should be read as a
supplement to those notes.
\end{abstract}

\vspace{1cm}
\begin{flushleft}
April 2004
\end{flushleft}
\end{titlepage}
\newpage


\section{Introduction}
\label{sec:introduction}

String field theory is a nonperturbative, off-shell formulation of
string theory  in target space.  Over the last several years,
nonperturbative solutions of string field theory have been identified
and studied in detail.  These solutions represent nonperturbative
shifts in the string background.  The fact that string field theory
admits such solutions is an important manifestation of the background
independence of the theory.  Developing a completely
background-independent formulation of string theory, where different
string backgrounds arise on equal footing as solutions of a system of
equations expressed in terms of a common set of degrees of freedom is
probably necessary in order to have a sensible description of
cosmology and early-universe physics in terms of string theory.
An introduction to string field theory and  recent developments in
this area
is given in the ``TASI 2001'' lecture notes by Taylor and Zwiebach
\cite{2001}; those lecture notes give a detailed introduction to
Witten's cubic string field theory and describe nonperturbative
results on tachyon condensation in this theory.  

While the primary motivation for string field theory is to develop
tools for understanding nonperturbative phenomena in string theory,
recent developments have also shown that string field theory can be
useful in performing perturbative off-shell and on-shell computations
in string theory.  These lecture notes give a short introduction to
such calculations.  They should be read as a supplement to
\cite{2001}.  Throughout these notes we use the notation and
conventions of \cite{2001}.  We assume that the reader is familiar in
particular with the material in sections 4 and 6 of \cite{2001}, where
the basic structure of Witten's OSFT is explained in detail.  The
content of these notes formed part of the material for lectures by W.\
Taylor at the TASI 2003 summer school; the remainder of the material
discussed in those lectures is described in \cite{2001}.
\vspace{0.2in}

The fact that Witten's cubic bosonic open string field theory
\cite{Witten-SFT} reproduces all on-shell perturbative string
amplitudes was demonstrated many years ago
\cite{Giddings-Martinec,gmw,Zwiebach-proof}.  In perturbative string
theory, a general on-shell $N$-point amplitude is computed by
performing an integral of the form
\[
\int {\rm d} \mu \; \langle{\cal O}_1 \cdots{\cal O}_n \rangle_\mu \,,
\]
where ${\cal O}_i$ are a set of vertex operators on the string world
sheet, and the integral $\int {\rm d} \mu$ is taken over the moduli
space of Riemann surfaces with $n$ marked points, summed over
topology.  In general, the modular integral becomes progressively more
difficult to perform explicitly as the topology of the Riemann surface
becomes more complicated.  Recent computations of genus 2 amplitudes
in superstring theory are described in \cite{dp}; analogous amplitudes
for genus 3 or higher have not yet been computed.

A nice feature of string field theory is that the integration over
moduli space is carried out in a natural and simple fashion.  In
Feynman-Siegel gauge, the SFT propagator is
\begin{equation}
\frac{1}{L_0}  = \int_0^{\infty}  {\rm d} t \; e^{-tL_0} \,.
\label{eq:propagator}
\end{equation}
Here, the Schwinger parameter $t$ can be interpreted in the string
world-sheet language as a modular parameter describing the length of a
propagating strip.  Thus, in SFT the modular integral simply becomes
an integration over a set of Schwinger parameters, which is quite
straightforward to organize.  For example, consider the planar
one-loop 3-point function for open strings.  The world-sheet in this
case has the form of an annulus with three marked points on the
boundary.  The usual parameterization of the world-sheet moduli space
would be in terms of the ratio of radii $r_2/r_1$ for the annulus and
the angles $\theta, \phi$ between the marked points.  In SFT, the
modular parameters are simply the lengths $t_i, i = 1, 2, 3$ of the
three strips connecting three 3-point vertices connected in a circle,
each with an external edge.  While the modular integrals in these two
coordinatizations can be related through a conformal map, the explicit
relationship between the Witten SFT parameterization and the usual CFT
parameterization becomes quite complicated for surfaces of higher
topology.  It was shown in \cite{Giddings-Martinec,gmw,Zwiebach-proof}
that in Feynman-Siegel gauge, the set of SFT diagrams with the
propagator (\ref{eq:propagator}) precisely reproduces all on-shell
open string amplitudes by covering the moduli space of Riemann
surfaces with the correct measure.  This guarantees that general
perturbative calculations in SFT will reproduce the usual open string
scattering amplitudes when restricted to on-shell external momenta.  An
example of this correspondence is the tree level 4-point Veneziano
amplitude, which was shown by Giddings \cite{Giddings} to be correctly
reproduced by Witten's SFT; we will consider this example in more
detail in Section 3.

While on-shell amplitudes are  invariant under arbitrary conformal
transformations of the string world-sheet, the same is not true of
off-shell amplitudes.  Generally, a different choice of conformal
coordinates will give a different definition for off-shell
amplitudes.  Essentially, string field theory provides a consistent
decomposition of the Riemann surface for each string diagram, giving a
systematic off-shell extension of string theory.  The goal of these
lectures is to describe methods for computing off-shell amplitudes
using Witten's open string field theory.


\section{Computing off-shell amplitudes}
\label{sec:off-shell}

There are a number of different ways of using OSFT to compute
off-shell amplitudes.  We briefly summarize four of these approaches
here:

\begin{enumerate}
\item[a)] Naive field theory approach:\\ In this approach, we treat
SFT as a standard field theory, albeit with an infinite number of
space-time fields.  We evaluate the amplitude by summing over all
fields which may arise on each internal propagator.  For example, the
tree level 4-point amplitude for the tachyon would be written
abstractly as
\begin{equation}
\sum_{I}^{}  V_{I \phi_1 \phi_2} \frac{1}{L_0 (I)}  V_{I \phi_3
  \phi_4} \,,
\end{equation}
where $I$ is the internal field, and $\phi_i$ are the external tachyon
fields.  Such computations cannot currently be performed exactly,
except in the simplest cases\footnote{for an example of an exact
  computation of this type, see \cite{Berkovits-Schnabl}, where the
off-shell zero-momentum  4-point function of the massless gauge field
is computed exactly}.  It is possible, however, to approximate such a
calculation by truncating the sum over intermediate fields at a fixed
mass level $L$.  We will refer to such an approximation as ``level
truncation on fields'', to distinguish it from a related method
described below involving level truncation on oscillators.  Because
the number of fields grows exponentially in $L$, the complexity of
this method grows exponentially in $L$.  This method was used in both
earlier \cite{ks-open} and more recent work
\cite{Sen-Zwiebach,Moeller-Taylor} on the tachyon condensation problem
in OSFT to compute coefficients in the tachyon effective potential.

\item[b)] Conformal mapping:\\ 
Another approach is to explicitly
construct a conformal mapping which takes the Witten parameterization
of moduli space to a parameterization more convenient for standard
conformal field theory calculations.  This is the method first used by
Giddings \cite{Giddings} to compute the Veneziano amplitude in
Witten's OSFT.  Following Giddings' calculation, this method was used
to compute the off-shell generalization of the Veneziano amplitude
\cite{Sloan,Samuel-off}, the off-shell 4-point function for general
fields \cite{Lechtenfeld-Samuel}, the off-shell Koba-Nielsen (tree
level N-point) amplitude \cite{Bluhm-Samuel-kn}, and the off-shell one-loop
N-point amplitude \cite{Bluhm-Samuel-1l}.  Finally, Samuel showed
\cite{Samuel-solving} that an arbitrary loop amplitude could be
computed using this approach.  This approach has the advantage that it
gives, in principle, analytic results.  These analytic results are often very
complicated, however, as the conformal mapping can be quite nontrivial
even in simple cases.

\item[c)] Oscillator method:\\
The key observation underlying the oscillator method is the fact that
both the cubic string vertex and the propagator can be expressed in
terms of exponentials of quadratic forms in the oscillator raising and
lowering operators.  Given a particular  diagram topology, with $v$
cubic vertices and $e$ internal edges, the set of cubic vertices
describes a squeezed state
\begin{equation}
\langle V | =\langle V_3 |_{123} \otimes
\langle V_3 |_{456} \otimes
\cdots  \otimes\langle V_3 |_{(3v-2) (3v-1) (3v)}  \in
({{\cal H}^*})^{3v}\,.
\label{eq:v}
\end{equation}
(The 2- and 3-vertices $| V_2 \rangle, | V_3 \rangle$ are described in
detail in section 6 of \cite{2001}.)
The $e$ internal edges connecting the labeled edges $i_k, j_k \leq 3v,
k \leq e$,
can be described by taking a squeezed state
\begin{equation}
| D \rangle = | V_2 \rangle_{i_1 j_1} \otimes
 | V_2 \rangle_{i_2 j_2} \otimes \cdots \otimes
 | V_2 \rangle_{i_e j_e} \,,
\end{equation}
describing the connection between the edges, and inserting a
propagator
\begin{equation}
P = \int \prod_{k = 1}^{e}  dT_k e^{-\frac{1}{2}T_k
  (L_0^{(i_k)}+L_0^{(j_k)})} \,.
\end{equation}
The complete set of amplitudes associated with the diagram of interest
is then given by
 an integral over internal (loop) momenta
\begin{equation}
{\cal A} = \int \prod_{i = 1}^{1 + e-v}  d^{26} q_i \;
\langle V | P | D \rangle\,.
\label{eq:amplitude}
\end{equation}
This expression gives a state in $({{\cal H}^*})^{3v-2e}$, which can
be contracted with external string states to get any particular
amplitude associated with the relevant diagram.  Because the
vertices are described by squeezed states, and $L_0$ is quadratic in
oscillator modes, the expression (\ref{eq:amplitude}) gives an
integral over squeezed states; the integrand can be computed using
standard squeezed state methods in terms of the infinite-dimensional
Neumann matrices $V^{rs}_{nm}, X^{rs}_{nm}$.  Thus, a
closed-form expression can be computed for any amplitude.  Using
current technology, such amplitudes cannot be computed exactly.
Truncating the Neumann matrices to finite size by imposing a cutoff on
level number leads to expressions in terms of finite-size matrices
which can be readily calculated.  The resulting amplitudes include
contributions from an infinite family of space-time fields, namely
those fields associated with string states containing oscillators up
to a fixed level $L$.  We refer to this method as ``level truncation
on oscillators''.  This method, unlike method (a), grows polynomially
in $L$.  This method was developed for Witten's OSFT in
\cite{WT-perturbative}.

\item[d)] Another method for computing OSFT amplitudes which has been
  developed recently, primarily by Bars, Kishimoto, Matsuo, and Park
  \cite{Bars-all} relies on the Moyal product representation of the
  star product \cite{Moyal}.  The idea here is that the star product
  can be diagonalized at the expense of complicating the operator
  $L_0$.  This method presents an alternative approach to performing
  calculations which seems to give numerical results comparable to
  those achieved using level truncation and the usual oscillator
  representation of the vertices and propagator as in methods (a),
  (c).  For a review of this approach see \cite{Bars}.

\end{enumerate}

In these lectures, we will describe some particular open string
amplitudes which have been computed using methods (a) and/or (c).
At this point we make some brief comments summarizing some relevant
features of the four methods just
described.

\begin{itemize}
\item Amplitudes computed using method (b)  are in principle
  exact, while methods (a, c, d) are approximate.  In all cases,
  however, numerical approximations are necessary to get concrete
  numbers for the amplitudes; in general, method (b) leads to complicated
  expressions in terms of quantities defined implicitly through
  integral relations, which can only be approximated numerically.
\item For finite amplitudes, the methods of field and oscillator level
  truncation (approaches (a, c)) have been found empirically to give
  errors which go as $a_1/L + a_2/L^2 + \cdots$, where $a_n$ are
  undetermined constants, and $L$ is the level of truncation.  There
  exists at this time no general proof of this result, but it seems
  fairly robust, and leads to highly accurate predictions for
  quantities which are known through other methods.
\item For genus $g > 1$, it is  difficult to get even approximate
  results using standard CFT methods.  In an impressive paper
  \cite{Samuel-solving}, Samuel
showed that in principle any loop amplitude can be
  computed using method (b); the calculations involved, while rather
  complex, are well-defined for amplitudes at any loop order.
Methods (a, c) have the advantage that they are
  not any more complicated conceptually for higher-loop amplitudes
  than for tree amplitudes, although they are more involved
  algebraically; it would be straightforward to  automate calculations
  using these methods for diagrams of arbitrary complexity.
\item Method (c) is generally useful for accurate computation of
  $n$-point amplitudes for small $n$, while method (a) is better for
  large $n$, since (a) scales polynomially in $n$ but exponentially in
  $L$ while (c) scales polynomially in $L$ but exponentially in $n$.
\end{itemize}


\section{Example: 4-tachyon amplitude}
\label{sec:example}

As a simple example of calculational approaches (a) and (c) we now
consider the off-shell 4-tachyon amplitude.  We first describe the
calculation of this amplitude in some detail at $p = 0$, where it
corresponds to the quartic term in the effective potential for the
tachyon zero-mode.  At the end of this section we briefly describe the
generalization of the calculation to nonzero momentum.  This
calculation is described in greater detail in \cite{WT-perturbative}.

Applying the general methodology described in (c) to the 4-tachyon
amplitude, we wish to consider the amplitude
\begin{equation}
{\cal A} = \frac{1}{2} g^2\int {\rm d}t
\left(\langle V_3 |_{123} \otimes
\langle V_3 |_{456}\right)
e^{-t/2 (L_0^{(3)} + L_0^{(4)})}
\left(| 0 \rangle_1 \otimes| 0 \rangle_2 \otimes | V_2 \rangle_{34}
\otimes| 0 \rangle_5 \otimes| 0 \rangle_6 \right).
\end{equation}
In oscillator language, this amplitude becomes
\begin{eqnarray}
\frac{1}{2}(\kappa g)^2\lefteqn{\int {\rm d}t \;e^{t}
\left[ \langle 0 |_{34}
\exp \left( -\frac{1}{2}a^{(3)}_nV^{11}_{nm} a^{(3)}_m-
\frac{1}{2}a^{(4)}_nV^{11}_{nm} a^{(4)}_m \right)
 e^{-N^{(3)}t} \right.
}\label{eq:amplitude1}\\ &  & 
\hspace{1in}
\left.
\times\exp \left( -(a^{(3)}_{n})^{\dagger}C_{nm}
(a^{(4)}_{m})^{\dagger}
 \right)
| 0 \rangle_{34} \times
     {\rm ghosts} \right]
\nonumber 
\end{eqnarray}
where $C_{nm} = (-1)^n \delta_{nm}$, and where we have used $N^{(3)} =
N^{ (4)}$ for the total matter oscillator number.  The ghost part of
the expression is practically identical in form, replacing $V \rightarrow X$
and replacing the matter oscillators $a_n, a_m$ by $c_n, b_m$.

The matter part of the matrix element in (\ref{eq:amplitude1}) can be
written as an inner product of two squeezed states
\begin{equation}
I_{{\rm matter}} =
\langle 0 | e^{-\frac{1}{2}a \cdot \tilde{V} \cdot a}
e^{-\frac{1}{2}a^{\dagger} \cdot S (t) \cdot a^{\dagger}}
| 0 \rangle
\label{eq:i1}
\end{equation}
where
\begin{equation}
\tilde{V} =
\left(\begin{array}{cc}
V^{11}_{nm} & 0\\
0 & V^{11}_{nm}
\end{array}\right)
\end{equation}
and
\begin{equation}
S (t) = \left(\begin{array}{cc}
0 & \delta_{mn} (-1)^m e^{-mt}\\
 \delta_{mn} (-1)^m e^{-mt} & 0
\end{array}\right)
\end{equation}
are expressed in terms of blocks, each of which is an
infinite-dimensional matrix with indices $n, m \in {\bf Z}$.

The matrix generalization \cite{Kostelecky-Potting} of standard
formulae for squeezed state inner products gives
\begin{equation}
I_{{\rm matter}} = \det \left( \identity -S (t) \cdot \tilde{V}
\right)^{-13} \,.
\end{equation}
A similar result holds for the ghosts.  Writing
\begin{equation}
x = e^{-t},
\label{eq:x}
\end{equation}
 the full
amplitude (\ref{eq:amplitude1}) can then be given as
\begin{equation}
{\cal A} = \frac{(\kappa g)^2}{ 2}  \int_0^1 \frac{dx}{x^2} 
\frac{\det \left( \identity +S (x) \cdot \tilde{X}\right)}{ \det
  \left(\identity -S (x) \cdot \tilde{V} \right)^{13}}  \,.
\label{eq:amplitude2}
\end{equation}
The expression (\ref{eq:amplitude2}) diverges as $x \rightarrow 0$.
This divergence arises from the intermediate tachyon state.  If we are
interested in computing the effective tachyon potential at zero
momentum,
\begin{equation}
V_{{\rm eff}} = -\frac{1}{2}\phi^2 + \frac{g \kappa}{ 3}  \phi^3 + c_4
(\frac{g \kappa}{ 3} )^2 \phi^4 + \cdots,
\label{eq:veff}
\end{equation}
then we wish to drop terms associated with the intermediate tachyon
field.  Terms in the ratio of determinants appearing in
(\ref{eq:amplitude2}) at order $x^n$ correspond to intermediate states
in the 4-tachyon tree amplitude at excitation level $L = n$.  Thus, 
dropping the intermediate tachyon term  should give
\begin{equation}
c_4 = -\frac{9}{2} \int_0^1 \frac{dx}{x^2} 
\left[
\frac{\det (1+S \cdot \tilde{X})}{\det (1-S \cdot \tilde{V})^{13}} -1
\right] \,. 
\label{eq:c4}
\end{equation}
The result (\ref{eq:c4}) contains infinite matrices;  an exact
result for $c_4$ cannot
currently be computed analytically from this expression.  We can,
however, use level truncation on oscillators to give an expression in
terms of finite-size matrices which can be computed exactly.  For
example, let us consider the simplest truncation, namely the
truncation to only fields composed of oscillators $a_1, b_1,$ and
$c_1$.  The matrices $S, \tilde{V}, \tilde{X}$ then become the $2
\times 2$ matrices
\begin{equation}
S =\left(\begin{array}{cc}
0 & -x\\
-x & 0
\end{array} \right),
\label{eq:s11}
\end{equation}
\begin{equation}
\tilde{V} =\left(\begin{array}{cc}
\frac{5}{27} &  0\\
0 & \frac{5}{27}
\end{array} \right),
\label{eq:n11}
\end{equation}
\begin{equation}
\tilde{X} =\left(\begin{array}{cc}
-\frac{11}{27} &  0\\
0 & -\frac{11}{27}
\end{array} \right),
\label{eq:x11}
\end{equation}
where we have used $ V^{11}_{11} = 5/27, X^{11}_{11} = -11/27$.  This
gives the resulting formula for the oscillator level 1 approximation
to $c_4$
\begin{equation}
c_4^{(1)} = -\frac{9}{2} \int_0^1 \frac{dx}{x^2} 
\left[\frac{1-\frac{121}{729} x^2}{ (1-\frac{25}{729} x^2)^{13}}  -1 \right]\,.
\label{eq:c4-l1}
\end{equation}
Expanding the integrand in a power series in $x$, we have
\begin{equation}
c_4 \approx - \frac{9}{2} \int_0^1 dx \;
\left[\frac{68}{243}  + \frac{650}{19683}  x^2 + \cdots \right]\,.
\label{eq:c4e}
\end{equation}
The first term in the expansion arises from the two intermediate fields at
level $L = 2$ ( $Ba_{-1}\cdot a_{-1}| 0 \rangle, \beta b_{-1} c_{-1}|
0 \rangle $).  The second term in the expansion arises from level $L =
4$ fields, etc..  The expression (\ref{eq:c4-l1}) can be integrated
exactly, leading to the oscillator level 1 approximation
\begin{equation}
c_4^{(1)} \approx -1.309 \,.
\end{equation}
Considering only the leading term in (\ref{eq:c4e}) amounts to
performing a level truncation on fields at level $2$, and gives
\begin{equation}
c_4^{[2]} = -34/27 \approx -1.259 \,.
\label{eq:c4-2}
\end{equation}
If we wish to include all intermediate fields to level $4$ (method
(a)), we would need to include oscillators up to level 3, and include
the first two terms in an expansion analogous to (\ref{eq:c4e}).  This
calculation is easy to carry out, and gives
\begin{equation}
c_4^{[4]}  \approx -1.472 \,.
\label{eq:c4-4}
\end{equation}
Using the empirical result that error in level truncation goes as
$1/L$, we can fit the two data points (\ref{eq:c4-2}, \ref{eq:c4-4})
through
\begin{equation}
c_4^{[n]} = \alpha + \beta/n
\end{equation}
giving
\begin{equation}
\alpha \approx -1.686 \,.
\end{equation}
To three decimal places, the exact value for $c_4$ is
\begin{equation}
c_4^{{\rm exact}}  \approx -1.742 \,.
\label{eq:c4-exact}
\end{equation}
Thus, the simple approximations given by truncating intermediate
fields at levels 2 and 4 are able to reproduce the correct answer to
within  approximately 3\%.  The coefficient $c_4$ was computed in
\cite{ks-open} using method (b), and in \cite{WT-perturbative}
using method (c) and oscillator truncation up to $L = 100$.  While no
closed-form expression for the result is known, numerical
approximations in both cases give (\ref{eq:c4-exact}) with error of
order $10^{-4}$.

We have now described in some detail the calculation of the off-shell
zero-momentum 4-tachyon amplitude in SFT.  This calculation can be
generalized in a straightforward fashion to compute all coefficients
$c_n$ in (\ref{eq:veff}); the results of this calculation and their
significance for the tachyon condensation problem are discussed in
\cite{2001}.  The calculation can also be generalized easily to
include nonzero external momenta for the interacting tachyon fields,
giving an off-shell generalization of the Veneziano amplitude.  We now
briefly review this generalization.

The on-shell 4-point tachyon amplitude for open bosonic strings is
perhaps the best-known perturbative calculation in string theory
\cite{Veneziano}.  To 
compute the on-shell amplitude at tree level for four tachyons with
momenta $p_1, \ldots, p_4$, one computes the disk amplitude with four
tachyon insertions on the boundary.  Three of the marked points on the
boundary can be moved to canonical points through a conformal map from
the disk to itself; the remaining marked point is a modular parameter
which must be integrated over to compute the full 4-point amplitude.
Often, the disk is mapped to the upper half-plane, and the fixed
points are taken to be 0, 1, and $\infty$, while the remaining point
$\xi$ is integrated over.  The resulting on-shell Veneziano amplitude is
given by
\begin{equation}
{\cal A}^{[V]}_4 (p_1, p_2, p_3, p_4) =
B ( -s-1, -t-1)
\label{eq:Veneziano}
\end{equation}
where $B (u, v)$ is the Euler beta function, and $s, t$ are the
Mandelstam variables
\begin{equation}
s = -(p_1 + p_2)^2, \;\;\;\;\;
t = -(p_2 + p_3)^2, \;\;\;\;\;
\end{equation}
The Euler beta function has the integral representation
\begin{equation}
B (u, v) = \int_0^1 d\xi \;  \xi^{u-1} (1-\xi)^{v-1}\,.
\label{eq:integral-beta}
\end{equation}
This integral representation of the Veneziano amplitude is convergent
when $u, v > 0$, so that $s, t < -1$.  For positive real $s, t$ it is
necessary to perform an analytic continuation to make sense of
(\ref{eq:integral-beta}) away from the poles at positive integers.

Using method (b) discussed in Section 2, Giddings reproduced the
Veneziano amplitude from OSFT \cite{Giddings}.  This calculation was
generalized by Sloan and Samuel \cite{Sloan,Samuel-off} to off-shell
momenta.  They found that for momenta not satisfying $p^2 = 1$, the
beta function integral (\ref{eq:integral-beta}) is replaced by
\begin{equation}
\int_0^1 d\xi \;  \xi^{u-1} (1-\xi)^{v-1}\left(\frac{1}{2} \kappa
(\xi) \right)^{\sum_{i}^{} (p_i^2 -1) }\,,
\label{eq:off-shell-Veneziano}
\end{equation}
where $\kappa (\xi)$ is a complicated integral of functions defined
implicitly by relations on elliptic integrals and the Jacobi zeta
function.  This function arises due to the complicated relationship
between the modular parameter $\xi$ and the Witten parameter $x$ which
appears in the OSFT calculation through (\ref{eq:x}).

An alternative approach to computing the off-shell 4-point tachyon
amplitude for generic $p$ is to use method (c) as above
\cite{Samuel-off,WT-perturbative}.  Including the momenta, the
calculation proceeds almost exactly as before, except that in the
vertices and propagator, the exponents containing quadratic forms in
the raising and lowering oscillators also contain terms of the form
$ap, a^{\dagger} p,$ and $p^2$.  The full amplitude can still be
computed using squeezed state methods, and the result (\ref{eq:amplitude2})
takes the form
\begin{equation}
\frac{(\kappa g)^2}{ 2}  \int_0^1 \frac{dx}{x^2} 
\frac{\det \left( \identity +S (x) \cdot \tilde{X}\right)}{ \det
  \left(\identity -S (x) \cdot \tilde{V} \right)^{13}} 
e^{-\frac{1}{2}p_iQ^{ij} (x)p_j}
 \,,
\label{eq:amplitude-p}
\end{equation}
where the quadratic form $Q^{ij} (t)$ can be expressed in terms of the
Neumann coefficient matrices $V^{rs}_{nm}$.  Using level truncation on
oscillators to approximate this amplitude numerically gives results
which reproduce both the on-shell amplitude (\ref{eq:Veneziano}) and
the off-shell amplitude (\ref{eq:off-shell-Veneziano}) to high
precision \cite{WT-perturbative}.  While this approach does not give
an analytic result like method (b), it is much easier to generalize to
more complicated diagrams.


\section{Applications}
\label{sec:applications}

In this section we briefly summarize two applications of perturbative
computations in OSFT.  First, we discuss the computation of the
effective action for the massless vector field on a D-brane.  Second,
we discuss loop computations in OSFT.

\subsection{Application: effective action for $A_\mu$}
\label{sec:example-a}

In \cite{2001}, we described the computation of the effective action
for the zero-momentum tachyon field $S_{\rm eff} (\phi)$.  As discussed in
the previous section, the coefficients $c_n$ in this action can be
computed by computing the $n$-point function for the tachyon and
omitting the tachyon field on all internal propagators; this amounts to
integrating out all other fields from the theory.  In a similar
way, we can compute an effective action for the massless vector field
$A_\mu$ on a D-brane, by integrating out all the massive string
fields, as well as the tachyon field.  We could also consider
integrating out the massive fields with $M^2 > 0$,
but not the tachyon, to find an
action for both $\phi$ and $A_\mu$---we comment further on this
possibility briefly below---but to compare with known on-shell
scattering amplitudes of $A_\mu$, it is perhaps most interesting to
consider the effective action for $A_\mu$ alone.

The effective action $S_{\rm eff} (A)$ which we compute in this
fashion is essentially an off-shell low-energy action for a system of
one (or many, if we include Chan-Paton factors) D-brane(s).  One
complication in computing this action is associated with the issue of
gauge invariance.  We would like an effective action which retains the
usual gauge invariance $\delta A_\mu = \partial_\mu \lambda$ (and its
nonabelian generalization).  The method we described above
for computing perturbative amplitudes, however, involved completely fixing
gauge to Feynman-Siegel gauge.  To retain some gauge invariance in the
effective action, we choose not to fix the gauge associated with the
field
\begin{equation}
 \lambda (p) b_{-1}| 0; p \rangle \,.
\label{eq:lambda}
\end{equation}
This leads to an extra space-time field associated with a state at level one,
\begin{equation}
\chi (p) b_{-1} c_0| 0; p \rangle,
\end{equation}
which must be included in the string field.  In the low-energy action,
$\chi$ becomes an auxiliary field which can  be explicitly
integrated out, giving a gauge-invariant action in terms of the vector
field $A_\mu$.

In the abelian case of a single D-brane, we know what to expect from
this calculation.  Because the on-shell condition for the vector field is
is $p^2 = 0$, we can perform a systematic double expansion of
the action in $p$ and $A$. The action should then take the form
\begin{equation}
S  \sim \sqrt{-\det (\eta_{\mu \nu} + F_{\mu \nu})} + {\rm derivative\
  corrections} 
\label{eq:abelian-BI}
\end{equation}
where $F_{\mu \nu}$ is the abelian Yang-Mills field strength and
the terms of order $ F^n \sim\partial^nA^n$ organize into the Born-Infeld
form, with terms of order $p^{n + k}A^n, k > 0$ considered as
derivative corrections.  In the nonabelian case, the situation is more
subtle.  Because derivatives can be exchanged with field strengths
through relations of the form $[D,[D, F]] =[F, F]$, we cannot expand
separately in $A, p$.  Instead, all terms at order $p^kA^{n-k}$ should
be considered together.  In this case we expect that the action will
take the ``nonabelian Born-Infeld'' form
\begin{equation}
S = -\frac{1}{4} {\rm Tr}\; F^2 + \alpha_3 {\rm Tr}\; F^3 +
\alpha_4{\rm STr}\; \left( F^4-\frac{1}{4} (F^2)^2 \right) + \cdots
\end{equation}
Much recent work has focused on computing the terms in this nonabelian
action in the case of the superstring; for some recent results and
further pointers to the literature see \cite{nonabelian}.  In the
bosonic case, the action has been computed to order $F^4$
\cite{Neveu-Scherk,Scherk-Schwarz, Tseytlin86}, and the
coefficients $\alpha_3, \alpha_4$ are known to be $\alpha_3 = 2ig_{\rm
  YM}/3, \alpha_4 = (2 \pi g_{\rm YM})^2/8$.

In \cite{Coletti-Sigalov-Taylor}, we computed all terms in the
effective action for the massless field $A$ to order $\partial^4A^4$
using a finite level expansion and extrapolation in a power series in
$1/L$ to compute all coefficients to 6 digits of accuracy.  This
computation correctly reproduces the Yang-Mills action both in the
abelian and nonabelian theories.  At finite level, the quartic term
$(A^2)^2$ in the bosonic theory does not vanish, due to terms in the
action of the form $\phi A^2$.  These terms, however, cancel to a high
degree of accuracy when higher-level fields are included.  It was
shown analytically in \cite{Berkovits-Schnabl} that this cancellation
is exact---in that paper, the authors also carried out the analogous
computation for the superstring, showing that the Yang-Mills action is
also reproduced correctly in that theory.

The first novel result of the explicit calculation of the effective
action for $A$ is the appearance of nonzero terms at order $p^2 A^4$
in the abelian theory.  Such terms do not appear in the Born-Infeld
action (\ref{eq:abelian-BI}).  The explanation for this discrepancy is
rather illuminating. In the full SFT, the transformation rule for
$A_\mu$ contains not only the term $\partial_\mu \lambda$ linear in
the gauge parameter appearing in (\ref{eq:lambda}), but also terms
which are linear in $\lambda$ and proportional to the massive string
fields.  When the massive fields are integrated out, the full
transformation rule becomes
\begin{equation}
\delta A_\mu = \partial_\mu \lambda +{\cal O} (\lambda A^2) \,.
\end{equation}
In order to relate the effective field transforming in this way to the
usual gauge field $\hat{A}_\mu$ transforming only under $\partial_\mu
\lambda$, it is necessary to perform a field redefinition
\begin{equation}
\hat{A}_\mu = A_\mu + \gamma_1 A^2 A_\mu + \gamma_2 A^2 \partial^2
A_\mu + \cdots \,.
\label{eq:redefinition}
\end{equation}
Field redefinitions of this type which relate SFT fields to the usual
space-time fields associated with CFT states were discussed in
\cite{Ghoshal-Sen,David}.  It was shown in
\cite{Coletti-Sigalov-Taylor} that after such a field redefinition,
the $F^3$ and $F^4$ terms in the nonabelian and abelian Born-Infeld
actions are correctly reproduced from SFT.  Some comments regarding
this computation may be helpful:
\begin{itemize}
\item It is important to emphasize that the ``natural'' Yang-Mills
variables appropriate for describing physics on a D-brane
are not at all apparent in SFT.  To get to the natural
fields for the D-brane background, it is necessary to do a field
redefinition.  On the one hand, this makes the SFT formulation of the
theory in the D-brane background seem quite obscure.  On the other
hand, this complication is a very natural feature of a
background-independent theory.  In any background-independent theory,
one should expect to have to do a complicated field redefinition to go
from the degrees of freedom associated with one background to the
degrees of freedom associated with another background.  We see here
that even though we have chosen a particular background around which
to expand the theory, SFT does not really use the natural fields
associated with that background.  This highlights the complexity of
describing physics in distinct backgrounds within the framework of SFT
or any other background-independent theory.  The gravitational
analogue of this observation is that in closed string field theory, or
any other formulation of string theory which is independent of closed
string backgrounds, we might expect that to write the usual metric
degrees of freedom associated with a particular background in terms of
the closed string fields, a field redefinition of the type
\begin{equation}
\hat{g}_{\mu \nu} = g_{\mu \nu} + \alpha g_{\mu \lambda}
g^{\lambda}_{\; \nu} + \cdots
\end{equation}
would be necessary

\item Another point worth noting is that the appearance of derivatives
  in the field
  redefinition (\ref{eq:redefinition})  has
  important implications for the geometrical interpretation of the
  effective theory.  In the usual Born-Infeld action, the transverse
  scalar fields $X^i$ encode the transverse positions of the
  D-branes.  On a D-brane of nonzero codimension, these fields are
  related through T-duality to the gauge field components $A_i$ on a
  higher-dimensional D-brane, and will transform under the same type
  of transformation rule as (\ref{eq:redefinition}).
Two D-branes with identical values of the scalars $\hat{X}^i$
at a
  particular point $\xi$ in the D-brane world-volume are coincident at
  that point, and have a massless string mode attaching them.  On the
  other hand, after a field redefinition of the form
  (\ref{eq:redefinition}), this geometry becomes much more obscure, as
  the redefined fields $X^i$ need {\it not} be the same even at a
  point of coincidence.  Thus, in the SFT formulation, the notion of
  locality on D-branes is rather subtly encoded in the fields of the theory.

\item The calculation described above could also be generalized to
  compute an effective action $S_{\rm eff}[\phi, A]$ for both the
  tachyon field $\phi$ and the gauge field $A_\mu$, as mentioned
  above,  by only integrating
  out fields with positive mass squared.  Because in this case,
  however, the on-shell equation for the tachyon is $p^2 = 1$ and not
  $p^2 = 0$, all derivative corrections will a priori be equally important, and
  there is no systematic expansion of this action in $p, \phi,$ and
  $A$.  Nonetheless, it may be that one can learn something of the
  physics of this theory by including terms to all orders in $p$.
\end{itemize}


\subsection{Application: Loops in OSFT}
\label{sec:loops}

As a final application of the methods described in Section 2, we
consider the computation of loop amplitudes.  Computing loop
amplitudes is necessary in order to ascertain whether OSFT is truly a
sensible quantum theory.  An extended discussion of these issues is
presented in the review article of Thorn \cite{Thorn-review}.  The
simplest loop diagrams are the annulus diagrams with one or two open
string vertices on the boundaries.  The one-loop two-point diagram was
computed by Freedman, Giddings, Shapiro and Thorn in \cite{fgst}, and
the one-loop one-point function was computed in
\cite{Ellwood-Shelton-Taylor}.

In \cite{fgst}, the one-loop two-point diagram was computed using
method (b) of Section 2.  These authors found good evidence that this
diagram has closed string poles, as expected.  Thus, if OSFT is to be
a good (unitary) quantum theory, it must in some way include closed
string degrees of freedom as asymptotic states.

In \cite{Ellwood-Shelton-Taylor}, we considered the one-loop open
string tadpole diagram, which we computed both using method (b) and
method (c).  Using both methods, it is possible to isolate the
divergences coming from closed strings.
Using method (c), the diagram can simply be written as
\begin{equation}
\langle T | = \int d^{26} q \int dt \langle V_3 |_{123}
e^{-tL_0^{(3)}} | V_2 \rangle_{23} \,,
\end{equation}
where $q$ is the internal (loop) momentum.  
This state in ${\cal H}^*$ gives the tadpole for any field in the
theory by contracting with the associated state in ${\cal H}$.  It can
be thought of as giving rise to a linear term in the effective action
$\langle T | \Phi \rangle$ where $| \Phi \rangle$ is the string field.
In terms of squeezed states, the tadpole $\langle T |$ can be written
as
\begin{equation}
\langle T | = \int_0^{\infty} {\rm d} t \; e^{t} \;
\langle 0 |c_0
\exp\left( -\frac{1}{2}a \cdot M (t) \cdot a-c \cdot R (t) \cdot
b\right) 
\frac{\det \left( \identity -S (t) \cdot \tilde{X}\right)}{ \left(Q (t)\det
  (
\identity -S (t) \cdot \tilde{V} )\right)^{13}} \,,
\label{eq:squeezed-tadpole}
\end{equation}
where $Q(t)$ is a scalar function of $t$ and
 $S, \tilde{X}, \tilde{V}, M, R$ are infinite matrices defined
appropriately for the one-loop tadpole diagram.
(The detailed form of these matrices is given in
\cite{Ellwood-Shelton-Taylor}).  This integral diverges as $t
\rightarrow 0$.  The appearance of this divergence is clear in the
closed string picture.  As $t \rightarrow 0$, the annulus becomes a
long, thin cylinder, which in the closed string picture represents a
closed string propagating along a world-sheet of length $s = 1/t
\rightarrow \infty$.  The closed string tachyon leads to a divergence,
as the closed string propagator takes the form
\begin{equation}
\int {\rm d} s \; e^{-sM^2} \,,
\end{equation}
which diverges when $M^2 < 0$ and $s \rightarrow \infty$.  This
divergence can be seen both in a computation using method (b) and in
(\ref{eq:squeezed-tadpole}).  In both cases, the form of the state 
$\langle T |$
as
$t \rightarrow 0$ corresponds to a Shapiro-Thorn state
\cite{Shapiro-Thorn} representing 
the closed string tachyon.

One very promising feature of this loop calculation is that we see
explicitly how the closed strings play a fundamental role in the open
string loop calculation.  The off-shell open string tadpole arises
essentially from the fact that in a D-brane background, the closed
strings acquire tadpoles associated with the linearized solution of
the supergravity equations in the D-brane background.  Because of the
coupling between open and closed strings, the closed string background
manifests as a nontrivial open string tadpole.  While this shows that
closed strings indeed arise naturally in OSFT, the difficulties
associated with the closed string tachyon seem to indicate that it is
unlikely that a complete and consistent formulation of the quantum
bosonic OSFT exists.

The tadpole (\ref{eq:squeezed-tadpole}) has in fact several types of
divergence.  There is a fairly harmless open string divergence
like that of (\ref{eq:amplitude2})
associated with the open string tachyon propagating around the loop in
the limit $t \rightarrow \infty$.  This divergence is easy to
analytically continue or remove by hand.  The divergence mentioned
above which is associated with the closed string tachyon, however, is
much more difficult to deal with systematically.  Removing this
divergence seems to require that we ``cheat'' by using our knowledge
of the closed string structure to analytically continue the closed
string tachyon.  This resolution is difficult to imagine continuing to
general loop order since the closed strings are not really
fundamental degrees of freedom in OSFT, so each diagram must
essentially be treated by hand, making any possibility of
nonperturbative results rather unlikely.  These problems seem to
indicate that, in the absence of some new and as-yet-unsuspected
magic, the quantum bosonic OSFT is probably not a well-defined unitary
theory, and that to have a complete quantum string field theory one
must probably
go to the superstring.  Indeed, there is as yet no reason to
believe that any of the problems encountered here should afflict the
superstring.  Several candidates for a classical superstring field
theory exist, including the Berkovits formalism \cite{Berkovits}, as
well as possibly a variation of the cubic Witten theory
\cite{Witten:1986qs,Wendt,Greensite-Klinkhamer,Arefeva:1989cp,Preitschopf:fc,Arefeva}---
for a recent comparison of these approaches see \cite{Ohmori-compare}.
A good classical superstring field theory should lead naturally to a
sensible quantum superstring field theory, at least in a background
associated with a D$p$-brane with $p < 7$.  For $p \geq 7$, there are
infrared divergences, also seen in the bosonic theory, which
complicate the problem somewhat; these divergences are simply those
which arise from the linearized gravity equations in the presence of a
source, and presumably do not represent a fundamental problem with the
theory.  It seems quite plausible that superstring field theory in a
D$p$-brane background can be formulated as a completely well-defined
and sensible nonperturbative quantum theory.


\section{Conclusions}
\label{sec:conclusions}

We have described several ways in which string field theory can be
used to compute on-shell and off-shell perturbative amplitudes.
String field theory offers a simple and consistent way to compute
arbitrary diagrams, including in principle even diagrams at very high
loop order.  On the other hand, at this point the methods available
for computing diagrams in SFT are primarily based on numerical
approximation methods.  These methods involve the truncation of an
infinite family of terms to a finite subset and an extrapolation by
matching to a simple power series in the reciprocal of the cutoff.

While we still do not have any completely rigorous demonstration of
the effectiveness of the level truncation method, empirical evidence
indicates that it converges well for a wide class of diagrams.  At
some point in the future it would be nice to develop a more systematic
approach to this approximation scheme, so that errors can be more
accurately estimated.  Still, even using fairly basic methods, we have
been able to calculate useful amplitudes to 5 or 6 digits of precision
using the level cutoff scheme.

In principle, OSFT coupled with level truncation can be used to
compute arbitrary string amplitudes to arbitrary precision.  For the
bosonic theory, however, loop amplitudes are plagued with various
divergences.  Among these, the divergences from the closed string
tachyon seem to render the theory sufficiently unstable that it is
difficult to imagine extracting physically useful information from
bosonic loop amplitudes, let alone nonperturbative quantum effects in
the bosonic theory.  To begin to address questions relevant to a
physical theory, it is probably necessary to have a consistent
formulation of superstring field theory.  Berkovits \cite{Berkovits}
has made some progress in this direction, and initial perturbative
results in the theory such as that of \cite{Berkovits-Schnabl} are
quite promising, but the quantum features of superstring field theory
remain to be investigated.

\section*{Acknowledgments}
I would like to thank the organizers and students from TASI 2003 for
creating a lively and stimulating atmosphere and for comments and
questions on the lectures.  I would also like to thank my
collaborators in the work described here: Erasmo Coletti, Ian Ellwood,
Nicolas Moeller, Jessie Shelton, and Ilya Sigalov.  Thanks to Barton
Zwiebach for numerous useful discussions and for collaborating on our
lecture notes \cite{2001} to which these notes form a supplement.
Thanks also to Olaf Lechtenfeld for discussions and correspondence
regarding some of the earlier work in this area.  This work was
supported by the DOE through contract \#DE-FC02-94ER40818.


\begin{thebibliography}{99}

\bibitem{2001}
W.~Taylor and B.~Zwiebach,
``D-branes, tachyons, and string field theory,''
{\tt arXiv:hep-th/0311017}.


\bibitem{Witten-SFT}
E.\ Witten, ``Non-commutative geometry and string field theory,''
\NP {\bf B268} 253, (1986).

\bibitem{Giddings-Martinec}
S.~B.~Giddings and E.~J.~Martinec,
``Conformal Geometry and String Field Theory,''
\NP {\bf B278}, 91 (1986).

\bibitem{gmw}
S.~B.~Giddings, E.~J.~Martinec and E.~Witten,
``Modular Invariance In String Field Theory,''
Phys.\ Lett.\ B {\bf 176}, 362 (1986).

\bibitem{Zwiebach-proof}
B.~Zwiebach,
``A Proof That Witten's Open String Theory Gives A Single Cover Of
Moduli Space,''
Commun.\ Math.\ Phys.\  {\bf 142}, 193 (1991).

\bibitem{dp}
E.~D'Hoker and D.~H.~Phong,
``Two-loop superstrings. I-IV,''
Phys.\ Lett.\ B {\bf 529}, 241 (2002),
{\tt hep-th/0110247};
Nucl.\ Phys.\ B {\bf 636}, 3 (2002),
{\tt hep-th/0110283};
Nucl.\ Phys.\ B {\bf 636}, 61 (2002),
{\tt hep-th/0111016};
{\tt hep-th/0111040}.

\bibitem{Giddings}
S.\ Giddings, ``The Veneziano amplitude from interacting string field
theory,'' \NP {\bf B278}
242 (1986).

\bibitem{Berkovits-Schnabl}
N.~Berkovits and M.~Schnabl,
``Yang-Mills action from open superstring field theory,''
JHEP {\bf 0309}, 022 (2003),
{\tt arXiv:hep-th/0307019}.


\bibitem{ks-open}
V.\ A.\ Kostelecky and S.\ Samuel, ``On a nonperturbative vacuum for the open
            bosonic string,'' \NP {\bf B336} (1990), 263-296.

\bibitem{Sen-Zwiebach}
A.~Sen and B.~Zwiebach,
``Tachyon condensation in string field theory,''
JHEP {\bf 0003}, 002 (2000),
{\tt hep-th/9912249}.


\bibitem{Moeller-Taylor}
N.~Moeller and W.~Taylor,
``Level truncation and the tachyon in open bosonic string field theory,''
Nucl.\ Phys.\ B {\bf 583}, 105 (2000),
{\tt hep-th/0002237}.


\bibitem{Sloan}
J.\ H.\ Sloan, `` The scattering amplitude for four off-shell tachyons
from functional integrals,'' \NP {\bf B302}
349 (1988).

\bibitem{Samuel-off}
S.\ Samuel, ``Covariant off-shell string amplitudes,'' \NP {\bf B308}
285 (1988).


\bibitem{Lechtenfeld-Samuel}
O.~Lechtenfeld and S.~Samuel,
``Covariant Off-Shell String Amplitudes With Auxiliary Fields,''
Nucl.\ Phys.\ B {\bf 308}, 361 (1988).

\bibitem{Bluhm-Samuel-kn}
R.~Bluhm and S.~Samuel,
``The Off-Shell Koba-Nielsen Formula,''
Nucl.\ Phys.\ B {\bf 323}, 337 (1989).

\bibitem{Bluhm-Samuel-1l}
R.~Bluhm and S.~Samuel,
``Off-Shell Conformal Field Theory At The One Loop Level,''
Nucl.\ Phys.\ B {\bf 325}, 275 (1989).

\bibitem{Samuel-solving}
S.~Samuel,
``Solving The Open Bosonic String In Perturbation Theory,''
Nucl.\ Phys.\ B {\bf 341}, 513 (1990).



\bibitem{WT-perturbative}
W.~Taylor,
``Perturbative diagrams in string field theory,''
{\tt hep-th/0207132}.



\bibitem{Bars-all}
I.~Bars and Y.~Matsuo,
``Associativity anomaly in string field theory,''
Phys.\ Rev.\ D {\bf 65}, 126006 (2002)
{\tt hep-th/0202030};
I.~Bars and Y.~Matsuo,
``Computing in string field theory using the Moyal star product,''
Phys.\ Rev.\ D {\bf 66}, 066003 (2002)
{\tt hep-th/0204260};
I.~Bars, I.~Kishimoto and Y.~Matsuo,
``String amplitudes from Moyal string field theory,''
Phys.\ Rev.\ D {\bf 67}, 066002 (2003)
{\tt hep-th/0211131};
I.~Bars, I.~Kishimoto and Y.~Matsuo,
``Fermionic ghosts in Moyal string field theory,''
JHEP {\bf 0307}, 027 (2003)
{\tt hep-th/0304005};
I.~Bars, I.~Kishimoto and Y.~Matsuo,
``Analytic study of nonperturbative solutions in open string field  theory,''
Phys.\ Rev.\ D {\bf 67}, 126007 (2003),
{\tt arXiv:hep-th/0302151};
I.~Bars and I.~Y.~Park,
``Improved off-shell scattering amplitudes in string field theory and new
computational methods,''
{\tt arXiv:hep-th/0311264}.


\bibitem{Moyal}
I.~Bars,
``Map of Witten's * to Moyal's *,''
Phys.\ Lett.\ B {\bf 517}, 436 (2001)
{\tt hep-th/0106157};
L.~Rastelli, A.~Sen and B.~Zwiebach,
``Star algebra spectroscopy,''
JHEP {\bf 0203}, 029 (2002)
{\tt hep-th/0111281};
M.~R.~Douglas, H.~Liu, G.~Moore and B.~Zwiebach,
``Open string star as a continuous Moyal product,''
JHEP {\bf 0204}, 022 (2002)
{\tt hep-th/0202087};
B.~Feng, Y.~H.~He and N.~Moeller,
``The spectrum of the Neumann matrix with zero modes,''
JHEP {\bf 0204}, 038 (2002)
{\tt hep-th/0202176}.



\bibitem{Bars}
I.~Bars,
``MSFT: Moyal star formulation of string field theory,''
{\tt hep-th/0211238}.

\bibitem{Kostelecky-Potting}
V.~A.~Kostelecky and R.~Potting,
``Analytical construction of a nonperturbative vacuum for the open
bosonic string,''
Phys.\ Rev.\ D {\bf 63}, 046007 (2001),
{\tt hep-th/0008252}.

\bibitem{Veneziano}
G.\ Veneziano, ``Construction of a crossing-symmetric, Regge-behaved
amplitude for linearly rising trajectories,'' {\rm Nuovo Cim.} {\bf
57A} 190 (1968).

\bibitem{nonabelian}
A.~Sevrin and A.~Wijns,
``Higher order terms in the non-Abelian D-brane effective action and  magnetic
background fields,''
JHEP {\bf 0308}, 059 (2003),
{\tt arXiv:hep-th/0306260}.


\bibitem{Neveu-Scherk}
A.\ Neveu and J.\ Scherk, ``Connection between Yang-Mills fields and
dual models'', 
\NP {\bf B36} (1972) 155-161;

\bibitem{Scherk-Schwarz}
 J.\ Scherk and J.H.\ Schwarz, 
``Dual models for non-hadrons'', \NP {\bf B81} (1974) 118-144.

\bibitem{Tseytlin86} A.~A.~Tseytlin,
"Vector field effective action in the open superstring theory",
\NP {\bf B276} (1986) 391; \NP{\bf B291} (1987) 876 (errata);

\bibitem{Coletti-Sigalov-Taylor}
E.~Coletti, I.~Sigalov and W.~Taylor,
``Abelian and nonabelian vector field effective actions from string
field  theory,''
JHEP {\bf 0309}, 050 (2003)
{\tt hep-th/0306041}.

\bibitem{Ghoshal-Sen}
D.~Ghoshal and A.~Sen,
``Gauge and general coordinate invariance in nonpolynomial closed string theory,''
Nucl.\ Phys.\ B {\bf 380}, 103 (1992), 
{\tt hep-th/9110038}.

\bibitem{David}
J.\  R.\ David, ``U(1) gauge invariance from open string field
theory'', {\em JHEP} {\bf 0010} (2000) 017, {\tt  hep-th/0005085}.


\bibitem{Thorn-review}
C.~B.~Thorn,
``String Field Theory,''
Phys.\ Rept.\  {\bf 175}, 1 (1989).

\bibitem{fgst}
D.~Z.~Freedman, S.~B.~Giddings, J.~A.~Shapiro and C.~B.~Thorn,
``The Nonplanar One Loop Amplitude In Witten's String Field Theory,''
Nucl.\ Phys.\ B {\bf 298}, 253 (1988).

\bibitem{Ellwood-Shelton-Taylor}
I.~Ellwood, J.~Shelton and W.~Taylor,
``Tadpoles and closed string backgrounds in open string field theory,''
JHEP {\bf 0307}, 059 (2003)
{\tt hep-th/0304259}.

\bibitem{Shapiro-Thorn}
J.~A.~Shapiro and C.~B.~Thorn,
``BRST invariant transitions between open and closed strings,''
\PR {\bf D36}    432 (1987);
``closed string-open string transitions in Witten's string field
theory,''
Phys.\ Lett.\ B {\bf  194},  43 (1987).

\bibitem{Berkovits}
N.~Berkovits,
``Super-Poincare invariant
                          superstring field theory,'' \NP
		  {\bf B450} (1995), 90,
                          {\tt hep-th/9503099};
``Review of open superstring field theory,''
{\tt hep-th/0105230};
``The Ramond sector of open superstring field theory,''
JHEP {\bf 0111}, 047 (2001),
{\tt hep-th/0109100}.

\bibitem{Witten:1986qs}
E.~Witten,
``Interacting Field Theory Of Open Superstrings,''
Nucl.\ Phys.\ B {\bf 276}, 291 (1986).

\bibitem{Wendt}
C.\ Wendt, ``Scattering amplitudes and contact interactions in Witten's
            superstring field theory,'' \NP {\bf B314} (1989) 209.

\bibitem{Greensite-Klinkhamer}
J.\ Greensite and F.\ R.\ Klinkhamer, ``Superstring amplitudes and contact
            interactions,'' \PL {\bf B304} (1988) 108.

\bibitem{Arefeva:1989cp}
I.~Y.~Arefeva, P.~B.~Medvedev and A.~P.~Zubarev,
``New Representation For String Field Solves The Consistence Problem For Open
Superstring Field,'' Nucl.\ Phys.\ B {\bf 341}, 464 (1990).

\bibitem{Preitschopf:fc}
C.~R.~Preitschopf, C.~B.~Thorn and S.~A.~Yost,
``Superstring Field Theory,''
Nucl.\ Phys.\ B {\bf 337}, 363 (1990).

\bibitem{Arefeva}
I.~Y.~Aref'eva, A.~S.~Koshelev, D.~M.~Belov and P.~B.~Medvedev,
``Tachyon condensation in cubic superstring field theory,''
Nucl.\ Phys.\ B {\bf 638}, 3 (2002),
{\tt arXiv:hep-th/0011117}.


\bibitem{Ohmori-compare}
K.~Ohmori,
``Level-expansion analysis in NS superstring field theory revisited,''
{\tt arXiv:hep-th/0305103}.


\end{thebibliography}
\end{document}